\documentclass[11pt]{amsart}

\usepackage{graphicx}
\usepackage{subcaption}

\usepackage{tikz} 
\usepackage{wrapfig}
\usetikzlibrary{arrows,shapes,positioning}
\usetikzlibrary{decorations.markings}
\tikzstyle arrowstyle=[scale=1]  
\tikzstyle directed=[postaction={decorate,decoration={markings,
    mark=at position .65 with {\arrow[arrowstyle]{stealth}}}}]
\tikzstyle reverse directed=[postaction={decorate,decoration={markings,
    mark=at position .65 with {\arrowreversed[arrowstyle]{stealth};}}}]
\usepackage{float} 
\usepackage[margin=1.0in]{geometry}
\usepackage[skip=2pt,font=scriptsize]{caption}

\usepackage{amsmath,amsfonts,amsthm,amssymb,graphics,color}

\usepackage{hyperref}   
\usepackage[utf8]{inputenc}
\usepackage[T1]{fontenc}
\usepackage{enumerate}

\newcommand{\black}[1]{\color{black}}
        \definecolor{pink}{rgb}{1,0,1}

\usepackage[abbrev]{amsrefs}

\theoremstyle{definition}

\numberwithin{equation}{section}

\title[An HIV-style strategy to fight new diseases]{Mathematics indicates that an HIV-style strategy could be applied to manage the coronavirus}

\author[Julie Rowlett]{Julie Rowlett} \address{Mathematics Department, Chalmers University and the University of Gothenburg, 41296, Gothenburg Sweden \\ Photo credit to Helmut R\"ub} \email{julie.rowlett@chalmers.se} 

\begin{document} 
\maketitle 
\section{Can we adapt strategies used to fight HIV to create strategies to fight the coronavirus?} 
We live with many viruses that have no vaccine and no cure.  One notable example is HIV.  Although many effective treatments have been developed, there is still neither a cure nor a vaccine for HIV.  Nonetheless, most people do not live in constant fear of HIV, in spite of the fact that it is a deadly and incurable virus.  How do we manage this?  How do you protect yourself from HIV?  

\begin{figure}
\centering
\includegraphics[width=0.8\textwidth]{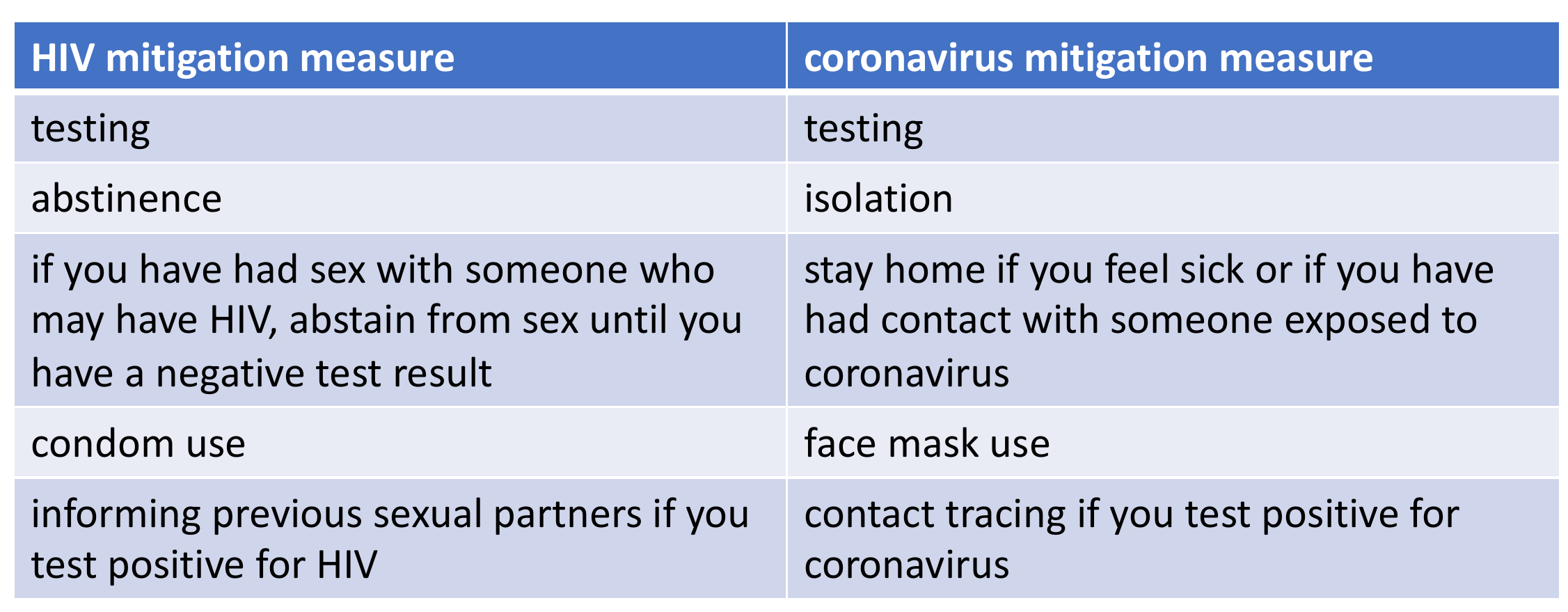}
\caption{There are many similarities between HIV and the coronavirus, like the fact that both viruses can be transmitted by people who show no symptoms.  To fight these invisible enemies, effective mitigation measures are also somewhat analogous.  Of course, the analogy is far from perfect because these viruses are also quite different.  For example, the level of intimacy required to contract HIV compared to the coronavirus is much greater.  Nonetheless, we may be able to apply lessons from fighting HIV to battle our new enemy.  These mitigation measure analogies are only a few; there may be further analogous measures that have escaped our attention.}
\label{fig:mm}
\end{figure}

You might answer that you abstain from sex with partners that have not been tested for HIV, or that you use condoms with new sexual partners.  These are examples of effective methods that when used correctly prevent or reduce transmission between people.  With the coronavirus and HIV, we highlight in Figure \ref{fig:mm} mitigation strategies for these two viruses that are somewhat - albeit not perfectly - analogous.  On the one hand, both HIV and the coronavirus can be transmitted by people who do not have any symptoms \cite{asymptomatic1, asymptomatic2, asymptomatic3, asymptomatic4}, so that both viruses can be invisible threats.  On the other hand, the transmission routes for HIV are much more specific and intimate compared with the transmission routes for the coronavirus.  Nonetheless, we may be able to use what we have learned in the past forty years fighting HIV and apply it to fight the novel coronavirus.  

The good news is:  we have very recently obtained a mathematical proof that an HIV-style strategy could work \cite{cjkjr2020}.  As with all theoretical mathematics, there are certain caveats that should be mentioned.  First, the mathematical model in \cite{cjkjr2020} is rooted in evolutionary game dynamics, that assumes individuals are rational and act in their best self-interest.  The model makes no predictions for individuals who do not fit that description.  Second, it is currently unknown whether or not infection from the coronavirus and subsequent recovery grants long-term immunity \cite{notimmune, notimmune2, notimmune3}.  Our model errs on the side of caution by making no assumption regarding long-term immunity; that is, we assume that immunity is either not conferred or is short-lived.  

\section{The disease dilemma: to mitigate or not to mitigate, that is the question} 
\begin{figure}
\centering
\includegraphics[width=0.8\textwidth]{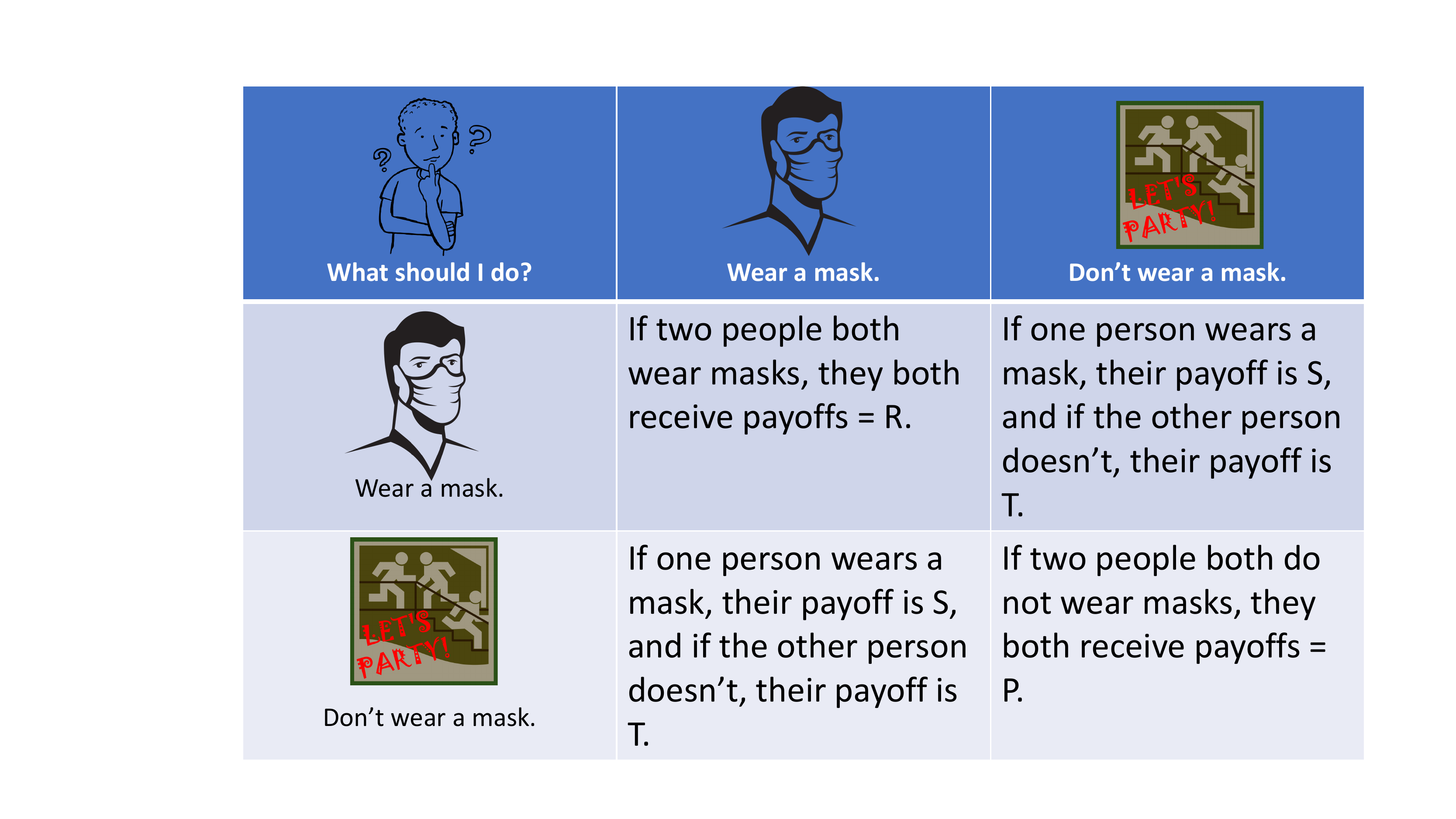}
\caption{In the `disease dilemma' people have the choice to cooperate, mitigating the spread of the disease, or defect, making no change to their regular behaviour.  This is described by the non-cooperative game shown here in normal form.  Image sources and license:  \href{https://openclipart.org/}{\texttt{openclipart.org}}, \href{https://creativecommons.org/publicdomain/zero/1.0/deed.en}{CC0 1.0}.}
\label{fig:dd}
\end{figure}

Our mathematical model combines the epidemiological model for the spread of diseases that do not grant lasting immunity together with the game theoretic model for predicting the evolution of human behaviors according to the replicator equation \cite{cjkjr2020}.   The epidemiological compartmental model is known as SIS (also known as SI) and has two compartments into which the population is categorized:  susceptible and infectious.  The classical SIS model does not incorporate human behavioral choices and changes, but human behavioral choices affect the spread of disease.  People can choose to change their usual behavior to include mitigation measures to reduce the transmission rate \cite{poletti2009}.  Moreover, people are not stuck with their choice; they are free to change their behaviors based on their perception of cost versus benefit. The World Health Organisation \cite{who} and numerous other references including \cite{ebola,asiatimes,stubborn} argue that it is reasonable to describe this situation with the Prisoner's Dilemma (PD) as depicted in Figure \ref{fig:dd}. 

In the context of coronavirus mitigation, research indicates that the most common transmission route is airborne \cite{jones2020, airbornetransmission,   airbornetransmission1, jama2020}, so that wearing a mask to prevent coronavirus transmission may be compared to wearing a condom to prevent the transmission of sexually transmitted infections.  We therefore focus on the mask mitigation measure.  Alice and Bob choose whether to cooperate, by wearing a mask, or defect, by not wearing a mask.  If Alice cooperates while Bob defects, then Alice pays the cost of buying the mask, which is represented by $-B<0$.  Alice receives some protection from her mask, represented by $\epsilon > 0$, but the main benefit is to everyone around Alice, similar to the reason surgeons wear masks not to protect themselves but to protect the person on the operating table \cite{masks}.  Consequently, if Bob does not wear a mask, he pays no cost but receives a benefit of $T> \epsilon > 0$.  Alice's total payoff is therefore $-B+\epsilon$.  Since the benefit to Alice is relatively small, we assume that $\epsilon < B$.  If both Alice and Bob cooperate, then they both pay the cost $-B$, but they also receive the maximal protective benefit of $T+\epsilon$, and their total payoffs are thus $T-B+\epsilon$.  Consequently, defining 
\[ C:= B - \epsilon,\] 
the payoffs satisfy 
\begin{equation} 
\label{pdpayoffs} 
S= -C < P= 0 <R< T = R+C. 
\end{equation} 

This particular representation of the Prisoner's Dilemma is known as the Donor-Recipient game and is given in normal form in Figure \ref{fig:dd}.  To generalize the two player game to model interactions within the entire population, the payoffs \eqref{pdpayoffs} are modified by a quantity $N(k)$, where $k$ is the average number of social contacts each individual in the society has.  Making a standard set of simplifying assumptions as in \cite{martin, nowak-graph, nowak-cycle, tanimoto-book, tanimoto1}, the payoffs $R$ and $P$ remain as in \eqref{pdpayoffs}, whereas the social network structure, that can be interpreted as peer pressure, now modifies the payoffs $S$ and $T$ 
\begin{equation} \label{pdpayoffs-network} 
S = - C + N(k), \quad T = R + C - N(k), \quad N(k) := \frac{Rk - 2C}{(k+1)(k-2)}, \quad k \in \mathbb N \setminus \{2 \}, \quad N(2) := R.
\end{equation}

\subsection{A dynamical system that combines disease spread with human behavior choices} 
The hybrid SIS-PD dynamical system we obtained in \cite{cjkjr2020} is 
\begin{align}  \label{eq:sispd} 
\dot I(t) &=\big([1-x(t)]\beta_D+x(t)\beta_C\big)I(t)(1-I(t))-\gamma I(t), \\
\dot x(t)&=x(t)(1-x(t))  [\alpha_1(\beta_D-\beta_C)I(t)-\alpha_2 (C-N(k))].  
\end{align} 
Above, the quantities on the left are differentiated with respect to time, $t$.  The frequency of cooperators in the population at time $t$ is $x(t)$, while the frequency of defectors is $1-x(t)$.  The positive quantities $\beta_C < \beta_D$ are the rates of transmission for cooperators and defectors, respectively.  The rate at which infected individuals become susceptible again is $\gamma$; if $D$ is the average duration of the infection, then $\gamma = 1/D$.  The rate of infectious individuals in the population is $I(t)$, and $1-I(t)$ is the rate of susceptible individuals in the population. 

There are three timescales in the model:  disease transmission, PD-payoff transmission, and information transmission.  The timescale of disease transmission is $t$, while the PD-payoff timescale is $\alpha_2^{-1} t$ with $\alpha_2 > 0$.  The timescale at which individuals receive disease-related information is $|\alpha_1|^{-1} t$.  The parameter $\alpha_1$ may be positive, negative, or zero and describes the frequency and accuracy of information disseminated about the disease.  Large values of $|\alpha_1|$ correspond to frequent exposure to information regarding the disease.  When $\alpha_1 > 0$, this corresponds to accurate information recommending disease avoidance, whereas when $\alpha_1<0$, this corresponds to (mis)-information which may suggest either the disease is harmless or that it is beneficial to contract the disease. For the sake of brevity, we refer to \cite{cjkjr2020} and related work by Poletti et. al \cite{poletti2009} for the derivation and justification of the system \eqref{eq:sispd} and its ability to accurately represent the spread of disease combined with the influence of human behavior choices.  

It is a straightforward exercise to calculate all equilibrium points of the hybrid dynamical SIS-PD system \eqref{eq:sispd}.  The asymptotically stable equilibrium points are summarized in Table \ref{table1}, with the key values of the information timescale parameter $\alpha_1$ below   
\begin{equation}\label{alpha1check} 
\check{\alpha}_1:= \frac{\beta_D}{\beta_D - \gamma} \frac{\alpha_2 (C-N(k))}{\beta_D - \beta_C}  \quad \mathrm{and}\quad \hat{\alpha}_1:= \frac{\beta_C}{\beta_C - \gamma} \frac{\alpha_2 (C-N(k))}{\beta_D - \beta_C}.
\end{equation}

\begin{table}[ht] 
	\centering
	\begin{tabular}{|l|l|}
		\hline
		\textbf{Range of $\alpha_1$} & \textbf{Unique stable equilibrium point for \eqref{eq:sispd} with $\alpha_1$ in this range}\\
		\hline
		$\qquad\ \alpha_1< \check{\alpha_1}$ & $x=0,\quad I= 1-\gamma/\beta_D$ \\
		\hline 
		$\check{\alpha_1} \leq \alpha_1\leq \hat{\alpha_1}$ & $x=x^*, \quad I= I^*$ \\
		\hline
		$\hat{\alpha_1} <\alpha_1$ & $x=1,\quad I = 1-\gamma/\beta_C$\\
		\hline
	\end{tabular}  
		\caption{\label{table1}These are all of the asymptotically stable equilibrium points of the SIS-PD model \eqref{eq:sispd}.  They depend on the value of $\alpha_1$ in relation to $\check{\alpha}_1$ and $\hat \alpha_1$ that are defined in \eqref{alpha1check}.  Above, $x^* = \frac{\beta_D}{\beta_D - \beta_C} - \frac{\gamma}{(\beta_D - \beta_C)(1-I^*)}$ and $ I^* = \frac{\alpha_2(C-N(k))}{\alpha_1(\beta_D - \beta_C)}$.}
\end{table}

\section{An HIV-style strategy to combat both the current and future pandemics} 
These results suggest a strategy to fight both diseases that do not to confer immunity as well as new diseases, since it is unknowable whether contracting and recovering from a \em new \em disease grants immunity \cite{mice}.  Mathematically, the strategy is to aim for the equilibrium point with $x=1$, and $I= 1-\gamma/\beta_C$, that is obtained if $\alpha_1$ is sufficiently large, corresponding to frequent reminders of effective mitigation measures.  In the limit, $x(t) \nearrow 1$, so the entire population of rational individuals acting in their best self interest cooperates.  As mitigation measures become increasingly effective, $\beta_C \searrow \gamma$, so that rate of infectious individuals $I(t)$ tends to zero.  Of course, as we learned in analysis, a limit may never actually be reached, but it can be approached.  The strategy we suggest, aiming for this limit, consists of two key steps.  

\textbf{Step 1:  }  Study the \em new virus \em to understand its particular features and thereby identify effective mitigation measures for  \em this particular new virus.  \em  Mathematically, the goal is to determine measures that minimize $\beta_C$, the rate of transmission among cooperators.  

\textbf{Step 2: }  Raise public awareness of the dangers of the disease caused by the virus, similar to the `Grim Reaper Ad Campaign' used in Australia to fight HIV \cite{hivaustralia2005, hivdownunder}.  In the case of covid-19, the disease caused by the coronavirus, healthy individuals are at risk of sustaining lung damage \cite{lungdamage}, brain and neural system damage \cite{braindamage, braindamage2, nervedamage}, organ damage \cite{lungandorgan}, sterility \cite{sterility}, or the worst outcome, death.  Conclude the advertisement with a positive and empowering message:  a clear explanation of the effective mitigation measures.  Mathematically, the ad campaign is used to increase the information transmission parameter $\alpha_1$, so that the unique stable equilibrium point in the dynamical system \eqref{eq:sispd} is that with $x=1$, and $I=1-\gamma/\beta_C$.  
 
\subsection{Concluding remarks: outsmarting viruses and evolving to a new normal.}  
We have learned to live with HIV and reduce the harm it causes by outsmarting the virus.  If the predictions of evolutionary game dynamics hold true, we can learn to live with coronavirus and reduce the harm it causes by outsmarting it with a strategy in the same spirit as the strategy we use to manage HIV.

\end{document}